\documentclass[jgrga]{agutex}
\authorrunninghead{HEYMAN ET AL.}
\titlerunninghead{TRANSPORT OF COARSE PARTICLES IN WATER}

\usepackage{graphicx}
\usepackage{amsfonts}
\usepackage{amsmath}
\usepackage{amssymb}
\usepackage{lineno}

\begin{document}

\title{Entrainment, motion and deposition of coarse particles transported by water over a sloping mobile bed}

\authors{J. Heyman\altaffilmark{1,3}, P. Bohorquez\altaffilmark{2} and C. Ancey\altaffilmark{3}}
\altaffiltext{3}{Laboratory of Environmental Hydraulics,  \'Ecole Polytechnique F\'ed\'erale de Lausanne, Switzerland}
\altaffiltext{2}{\'Area de Mec\'anica de Fluidos, Departamento de Ingenier\'ia Mec\'anica y Minera, CEACTierra, Universidad de Ja\'en, Campus de las Lagunillas, 23071, Ja\'en, Spain}
\altaffiltext{1}{Groupe Milieux Divis\'es, UMR CNRS 6626, Institut de Physique de Rennes, Universit\'e Rennes I, Campus de Beaulieu, F-35042 Rennes, France}

\authoraddr{Corresponding author: J. Heyman, UMR CNRS 6626, Institut de Physique de Rennes, 35042 Rennes, France (joris.heyman@univ-rennes1.fr)}

\begin{abstract}
In gravel-bed rivers, bedload transport exhibits considerable variability in time and space. Recently, stochastic bedload transport theories have been developed to address the mechanisms and effects of bedload transport fluctuations. Stochastic models involve parameters such as particle diffusivity, entrainment and deposition rates. The lack of hard information on how these parameters vary with flow conditions is a clear impediment to their application to real-world scenarios. In this paper, we determined the closure equations for the above parameters from laboratory experiments. We focused on shallow supercritical flow on a sloping mobile bed in straight channels, a setting that was representative of flow conditions in mountain rivers. Experiments were run at low sediment transport rates under steady nonuniform flow conditions (i.e., the water discharge was kept constant, but bedforms developed and migrated upstream, making flow nonuniform). Using image processing, we reconstructed particle paths to deduce the particle velocity and its probability distribution, particle diffusivity, and rates of deposition and entrainment.
We found that on average, particle acceleration, velocity and deposition rate were responsive to local flow conditions, whereas entrainment rate depended strongly on local bed activity. Particle diffusivity varied linearly with the depth-averaged flow velocity. The empirical probability distribution of particle velocity was well approximated by a Gaussian distribution when all particle positions were considered together. In contrast, the particles located in close vicinity to the bed had exponentially distributed velocities. Our experimental results provide closure equations for stochastic or deterministic bedload transport models.
\end{abstract}

\begin{article}

\section{Introduction}
Sediment transport has been studied since the earliest developments of hydraulics in the 19th century \citep{duboys79,gilbert14}. Despite research efforts, its quantification remains a notoriously thorny problem. This holds especially true for gravel-bed rivers, where multiple processes can interact with each other, making it difficult to predict sediment transport rates \citep{recking12b,recking13b}. A typical example is provided by how particle size distribution influences particle mobility, grain sorting, bed armouring, bed forms, varying flow resistance, bed porosity and hyporheic flow and eventually altering transport capacity in a complex way \citep{gomez91,church06,comiti12,powell14,yager15,rickenmann16}.

Quantification of sediment transport has long been underpinned by several key concepts. For instance, as sediment motion is driven by water flow, the sediment transport rate $q_s$ is routinely considered a one-to-one function of the water flow rate $q_w$, with a parametric dependence on particle size and bed slope. Numerous bedload transport equations in the algebraic form $q_s=f(q_w)$ have been developed from field and experimental data, mainly under the assumption of bed equilibrium conditions (i.e., bed slope is constant on average) \citep{garcia07,rickenmann16}. Applied to natural waterways, these equations perform poorly at predicting the sediment transport rate to within less than one order of magnitude \citep{Gomez1989,barry04,recking12b,ancey14ercoftac,gaeuman15}. The mere existence of several algebraic equations of similar structure combined with substantial data spread is a hint that something goes amiss in this approach. Various arguments have been used to explain this poor performance, such as the strong nonlinearity in the coupling between hydraulic variables, stress distribution, and sediment transport \citep{recking13b}, limited sediment availability \citep{lisle02,wainwright15}, partial grain mobility \citep{parker82b,wilcock97c}, hysteretic behavior under cyclic flow conditions \citep{wong06a,humphries12,mao12} and bed form migration  \citep{gomez91}.

Reasons for poor performance have also been investigated through detailed experiments. Strikingly, even under ideal flow conditions in the laboratory (i.e., bed equilibrium, steady uniform flow, initially planar bed, spherical particles of the same size, constant sediment supply and water discharge), the sediment transport rate $q_s$ exhibits large fluctuations \citep{boehm04,ancey08a}. This suggests that fluctuations are intrinsic to sediment transport, and they may be amplified under natural flow conditions as a result of bed form migration \citep{Whiting1988,gomez91} or partial fractional transport \citep{kuhnle88}. Further experimental investigations have revealed other remarkable features of $q_s$ time series such as intermittency, long spatial and temporal correlation, and multifractality \citep{singh09,radice09,radice09b,singh10,kuai12,campagnol12,heyman14}.

These experiments have created greater awareness of two fundamental aspects of bedload transport: randomness and the discrete nature of particle transport. As a consequence, a number of technical questions have been raised as to how the sediment transport rate should be defined theoretically and measured experimentally \citep{ancey10a,furbish12a,ballio14}. A lack of consensus has led to a rekindling of the debate about the physics underlying sediment transport, a debate initiated decades ago by \citet{einstein50} and \citet{bagnold66b}, among others. Both Einstein and Bagnold considered sediment transport at the particle scale, but with different assumptions: Einstein viewed particle flux as the imbalance between the number of particles entrained and then deposited on the bed, while Bagnold treated bedload transport as a two-phase flow whose dynamics are controlled by the momentum transferred between the water and solid phases.

In recent years, there has been renewed interest in grain-scale analysis of bedload transport, with an emphasis placed on the stochastic nature of particle motion. Two new families of stochastic models have emerged from Einstein's seminal work, while others have elaborated on Bagnold's ideas \citep{seminara02,lajeunesse10}. The first family follows the Lagrangian framework, in which particles are tracked individually. To deduce the bulk properties, such as particle flux and activity (i.e., the number of moving particles per unit streambed area), recent studies have focused on the statistical properties of particle trajectories in their random walks  \citep{ganti10,furbish12a,furbish12b,armanini14b,pelosi16,fan16}. An alternative is the Eulerian framework, which derives the bulk properties by averaging particle behavior over a control volume \citep{ancey08a,ancey14}.

Predictions from stochastic models have been successfully compared with experimental data in the laboratory, mostly under steady state conditions \citep{ancey08a,roseberry12a,hill10,heyman13,heyman14,fathel15}. The next step, comparison with field data, is much more demanding. Indeed, in real-world scenarios, water flow is rarely uniform as a result of the complex interplay between bed morphology, hydrodynamics, and sediment transport. As a consequence, stochastic models must be coupled with governing equations for the water phase (e.g., the shallow water equations), a task that raises numerous theoretical and computational problems \citep{bohorquez15,audusse15}. One of these problems is that existing stochastic models introduce a number of parameters (e.g., the particle diffusivity, the entrainment and deposition rates) without specifying how they depend on water flow. The absence of closure equations for stochastic models prevents their wider applicability. Furthermore, while the two families of stochastic models show consistency with each other, points of contention have also emerged and are not settled to date. A typical example is provided by the probability distribution of particle velocity, a key element in understanding particle diffusion. Authors have found that an exponential distribution fits their data well \citep{lajeunesse10,furbish13a,fan14,furbish16}, while others lean towards a Gaussian distribution \citep{martin12,ancey14}.

This paper aims to provide some of the equations needed to close stochastic models. Here we focus on shallow supercritical flows on sloping beds at low sediment transport rates. On the one hand, the experimental setting bears similarity with flow conditions encountered in mountain streams: a low submersion ratio, high water speeds, average bed slope in excess of 1\%, migrating bedforms, and coarse particles rolling or saltating along the bed. On the other hand, these conditions facilitate experimental analysis: low transport rates imply low particle velocities, irregular trajectories, and random deposition and entrainment events.

\section{Theoretical Background}
Bedload transport theory aims to calculate macroscopic quantities such as the bedload transport rate $q_s$, and the entrainment and deposition rates, $E$ and $D$, respectively. Stochastic theories follow the same objective, but they are based on the assumption that the bulk quantities $q_s$, $E$, and $D$ reflect the random behavior of particles on the microscopic scale, and so are driven by noise to a large extent. An insightful analogy can be drawn with turbulence: in the Reynolds-averaged Navier-Stokes equations, the turbulent stress tensor $-\langle\varrho{\boldsymbol u}'{\boldsymbol u}'\rangle$ results from the interactions between the fluctuating velocity components ${\boldsymbol u}'$  (with $\varrho$ fluid density and $\langle\cdot\rangle$ the ensemble average). One challenge in turbulence is to close the Reynolds-averaged Navier-Stokes equations by relating the Reynolds tensor $-\langle\varrho{\boldsymbol u}'{\boldsymbol u}'\rangle$ to the average velocity gradient. Stochastic bedload transport theory faces similar issues, and this is what we outline below.

If transported sediment behaved like a continuum, it would be natural to define the bedload transport rate as the particle flux across a control surface $S$
\begin{linenomath*}\begin{equation}
q_s=\int_S c {\boldsymbol v}_p\cdot{\boldsymbol n}{\text{d}} S
\label{eq:flux-mmc}
\end{equation}\end{linenomath*}
where ${\boldsymbol n}$ denotes the outward oriented normal to $S$, ${\boldsymbol v}_p=(u_p, v_p)$ is the particle velocity, and $c$ denotes the particle concentration. Yet, bedload transport involves disperse discrete particles, which implies that both particle concentration and velocity vary with time even under steady state conditions. Introducing a Reynolds-like decomposition ${\boldsymbol v}_p=\langle{\boldsymbol v}_p\rangle+{\boldsymbol v}_p'$ and $c=\langle c\rangle +c'$ leads to the definition of an ensemble averaged particle flux in the streamwise direction
\begin{linenomath*}\begin{equation}
\langle q_s\rangle = S\langle c\rangle\langle u_p\rangle  +S\langle c' u'_p\rangle
\label{eq:flux0}
\end{equation}\end{linenomath*}
As the velocity and concentration fluctuations are large, especially at low sediment transport rates \citep{cudden03,ancey08a,singh09,campagnol12,campagnol15}, the contribution $S\langle c' u'_p\rangle$ arising from fluctuations cannot be ignored. The crux of the problem is determining this contribution as a function of the average flow variables.

There are many different ways of looking at this issue \citep{ballio14}, and here we will only scratch the surface. An interesting outcome of recent developments in stochastic bedload load theory is the definition of the transport rate
\begin{linenomath*}\begin{equation}
\langle q_s\rangle =  V_s \left( \left \langle u_p\right \rangle \langle\gamma\rangle - \frac{\partial}{\partial x}(D_u\langle\gamma\rangle ) \right)
\label{eq:flux}
\end{equation}\end{linenomath*}
where $V_s$ is the particle volume, $D_u$ denotes particle diffusivity and $\gamma$ particle activity (i.e., the number of moving particles per unit streambed area). Comparison with Eq.~\eqref{eq:flux0} shows that $ V_s \langle\gamma\rangle=S\langle c\rangle$. Remarkably, equation \eqref{eq:flux} has been derived using Lagrangian \citep{furbish12a} and Eulerian \citep{ancey14} approaches. This definition of  bedload transport rate shows that the contribution due to fluctuations $S\langle c' u'_p\rangle$ is related to the streamwise gradient of particle activity via a parameter called the \emph{particle diffusivity} and thus provides a closure equation: $S\langle c' u'_p\rangle=V_s \partial_x(D_u\langle\gamma\rangle)$. The definition \eqref{eq:flux} of bedload transport rate involves three variables that we now have to specify: $\langle\gamma\rangle$, $\langle u_p\rangle$, and $D_u$.

Mass conservation implies that the time variations in particle activity are described by an advection-diffusion equation with a source term \citep{furbish12b,ancey14,ancey15jgr}
\begin{linenomath*}\begin{equation}
 \frac{\partial \langle\gamma\rangle}{\partial t}+  \frac{\partial  }{\partial x}(\langle u_p\rangle\langle\gamma\rangle)-\frac{\partial^2  }{\partial x^2}(D_u\langle\gamma\rangle) = E -D
\label{eq:gamma0}
\end{equation}\end{linenomath*}
which can also be cast in the following form \citep{charru04,lajeunesse10}
\begin{linenomath*}\begin{equation}
 \frac{\partial \langle\gamma\rangle}{\partial t}+ \frac{1}{V_s} \frac{\partial \langle q_s\rangle}{\partial x} = E-D
\label{eq:gamma}
\end{equation}\end{linenomath*}
where $E$ and $D$ denote the entrainment and deposition rates, i.e. the volume of sediment entrained and deposited per unit streambed area and per unit time. Equation \eqref{eq:gamma0} involves two other unknown functions, $E$ and $D$, to be determined. There is a dearth of hard information on how these rates are related to flow conditions. Two experiments suggest that $D$ is proportional  to $\langle\gamma\rangle$: $D=D_p\langle\gamma\rangle $, where the particle deposition rate $D_p$ was found to be roughly independent of bottom shear stress \citep{lajeunesse10,ancey08a}. The observations are less clear for the entrainment rate $E$, which was found either to increase linearly with shear stress \citep{lajeunesse10} or to be roughly independent of it \citep{ancey08a}.

\begin{figure}
\includegraphics[width=8cm]{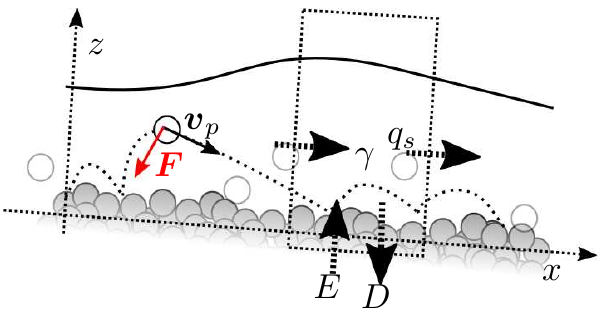}
\caption{Schematic representation of the transport process. The $x$-axis points down the flume, whereas $z$ is in the direction of the upward-pointing normal to the bed. The particle flux is denoted by $q_s$, $\gamma$ is the particle activity, $E$ and $D$ are the areal entrainment and deposition rates, $\boldsymbol{v}_p=(u_p,v_p)$ is the particle velocity, $\boldsymbol{F}$ the total force acting on the particle. }
\label{fig:theory}
\end{figure}

Average particle velocity has been extensively studied. Experiments have shown that  $\langle u_p\rangle$ is proportional to the shear velocity $u_*=\sqrt{\tau_b/\varrho}$
\begin{linenomath*}\begin{equation}
\langle u_p\rangle = A(u_*-u_c)
\label{eq:up-us}
\end{equation}\end{linenomath*}
where $u_c$ is a critical velocity corresponding to incipient motion and $A$ is a parameter ranging from 3 to 15 (sometimes 40) depending on flow conditions and bed features \citep{francis73,abbott77,vanrijn84a,nino94a,ancey02b, lajeunesse10,martin12}. Particle dynamics have also been investigated by taking a closer look at the time variation in the particle momentum
\begin{linenomath*}\begin{equation}
m_p \frac{{\text{d}} \boldsymbol{v}_p}{{\text{d}} t}=\boldsymbol{F}(t,\boldsymbol{v}_p,u_*, \theta, \cdots) \label{eq:motion2}
\end{equation}\end{linenomath*}
where $m_p$ is particle mass, $\boldsymbol{F}$ is the total force exerted on the grain at time $t$: hydrodynamic forces including pressure, drag and lift forces (possibly, other forces such as the added mass force) and contact forces due to friction and collision with the bed (see Fig.~\ref{fig:theory}). The total force is expected to depend on many variables, the most significant being the particle velocity, bed slope, and shear velocity. Of particular interest is the statistical behavior of the fluctuating part of $\boldsymbol{F}$ as it affects the shape of the probability distribution of particle velocities $P(u_p)$ and particle diffusivity $D_u$.

Particle diffusion is a direct consequence of the fluctuating force $\boldsymbol{F}'$. In the absence of fluctuations, particles move at the same velocity and they do not disperse. Note that the picture is blurred by the intermittent nature of bedload transport: even in the absence of particle dispersal, particle activity may exhibit a length-scale-dependent pseudo-diffusive behavior resulting from entrainment and deposition of particles \citep{ancey15jgr,campagnol15}. Here, for the sake of simplicity, we consider that particle diffusivity is a measure of particle displacement over time. Let us track one particle in its displacement and call $X(t)$ its position at time $t$. In the absence of force fluctuations, the particle's mean-square displacement $\langle X(t)^2\rangle$ varies as $t^2$ (ballistic regime): $\langle X(t)^2\rangle=\langle u_p^2\rangle t^2$. When particle motion is affected by fluctuations, $\langle X(t)^2\rangle$ exhibits a power-law scaling:  $\langle X(t)^2\rangle\propto t^n$. The case $n=1 $ corresponds to (normal) diffusion, and in that case the coefficient of proportionality is $2D_u$. The case $n> 1 $ is referred to as superdiffusion, while $n<1 $ is associated with the subdiffusive regime. Depending on the timescale of observation, bedload transport shows normal or abnormal diffusion \citep{nikora02,ganti10,zhang12,hassan13,phillips13,pelosi14,campagnol15}.

The calculation of the probability distribution of particle velocities $P(u_p)$ and particle diffusivity $D_u$ is a major challenge that is attracting growing attention. \citet{furbish13a} and \citet{furbish16} borrowed arguments from statistical mechanics to show that $P(u_p)$ was an exponential distribution, a result in close agreement with observations \citep{lajeunesse10,roseberry12a}.
Making an analogy with Brownian motion in a potential well, \citet{ancey14} assumed that $\boldsymbol{F}'$ behaves like white noise and $\langle\boldsymbol{F}\rangle $ relaxes to its steady state value over a certain characteristic time, and thereby they deduced that $P(u_p)$ was a truncated Gaussian distribution, a form supported by experimental evidence \citep{martin12}. Assuming that $\langle\boldsymbol{F}\rangle $ was constant and particles move sporadically, \citet{fan16} found that particles may exhibit subdiffusive, normal, or superdiffusive behavior depending on the resting time. While particle diffusion and more specifically the determination of particle diffusivity have been addressed experimentally \citep{heyman14,seizilles14}, there is scarce information on how $D_u$ varies with the flow conditions.

The objective of this article is to document the dependence of   $\langle u_p\rangle$, $E$, $D$, $D_u$ on flow conditions. Focus is given to shallow supercritical flows, and our results may differ from earlier results obtained under other flow conditions \citep{lajeunesse10,roseberry12a,seizilles14,fathel15}.

\section{Methods}
\subsection{Experimental Setup}
\begin{figure}[h]
\includegraphics[width=8cm]{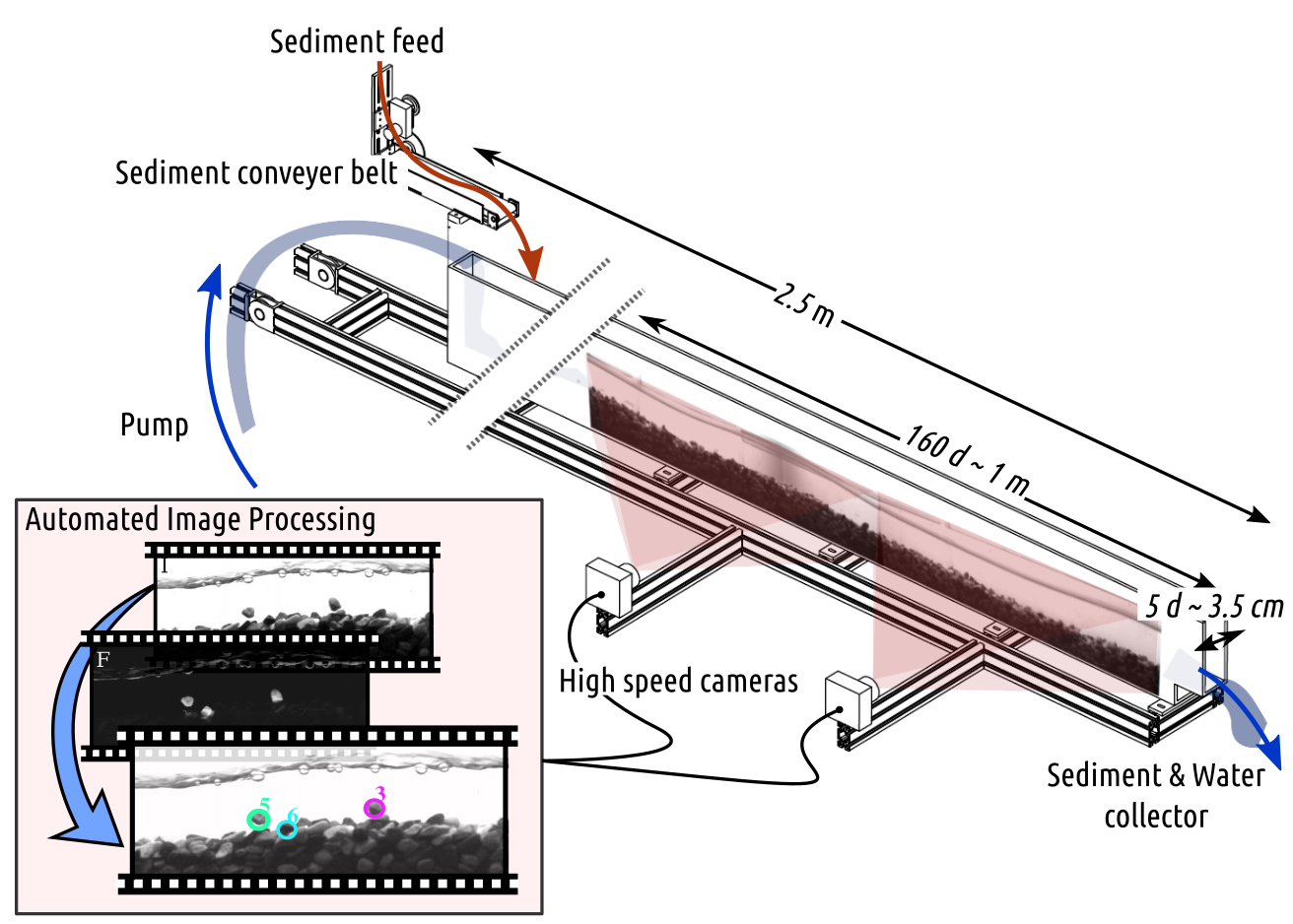}
\caption{Experimental setup ($d$ is the mean grain diameter).}
\label{fig:setup}
\end{figure}

Experiments were carried out in a 2.5-m long   3.5-cm wide flume  (see Fig.~\ref{fig:setup}). The water discharge was controlled using an electromagnetic flow meter. A hopper coupled to a conveyor belt fed the flume with sediment at a prescribed rate. Bed slope ranged from 3\% to 5\% on average (due to bedforms, local slope ranging from --30\% to 20\% was observed). Flow was shallow and supercritical. Sediment transport was low, with a Shields number ranging from 0.09 to 0.12 (see Table \ref{tab:adim}).

The bed was made up of natural gravel with a narrow grain size distribution: the mean diameter was $d=6.4$~mm, while the 30th and 90th percentiles were $6.1$~mm and $6.7$~mm, respectively. Gravel with a narrow grain size distribution had the advantage of facilitating particle tracking and limiting grain sorting.

Note that the flume was narrow (with a width of approximately 5 grain diameters). This configuration had many advantages over wider flumes. Among other things, it favored the formation of two-dimensional bedforms, whereas wider flumes are prone to form alternate bars. Furthermore, it made it possible to take sharp images from the sidewall. A disadvantage was the significant increase in flow resistance, possibly with a change in turbulent structures (that might be controlled by the sidewalls rather than bed roughness). Moreover, the confining pressure exerted by close sidewalls increases bulk friction within granular beds \citep{taberlet04}, which in turns affects bed stability and sediment transport \citep{Aalto1997,Zimmermann2010(2)}. In order to limit the bias induced by flume narrowness, we paid special attention to computing the bottom shear stress with due account taken of the sidewall influence (see section 3 in the Supporting Information).

Two digital cameras were placed side by side at the downstream end of the channel. They took $1280 \times 200$-pixel (px) pictures through the transparent sidewall at a rate of 200~frames per second (fps). The field of view in the streamwise direction covered approximately 1~m or (i.e., 160 grain diameters). In other word, image resolution was close to 0.4~mm/px (i.e., 16 pixels per grain). This length of the observation window was a tradeoff between highest resolution (to track particles) and longest distance that particles could travel.

We ran 10 experiments by varying mean bed slope, water discharge, and mean sediment transport rate (see Table S1 in the Supporting Information). For each experiment, we waited a few hours until that the flow and sediment transport rates reached steady state. Then, we filmed 2 to 8 sequences of 150~s at 200~fps (i.e., 30,000 frames). This corresponded to the maximum random access memory available on the computer used (30~GB). In some experiments, we acquired more sequences of 4,000 frames (i.e., 20~s).

\subsection{Image Processing}
Image processing (automatic particle tracking) was subsequently performed on the video frames using the following procedure. For each video frame, moving zones (consisting of moving particles) were detected using a background subtraction method. In this method, a typical ``background'' image showing bed particles at rest was built iteratively and subtracted from the current frame to obtain the ``foreground'' moving zones of the image (see the Supporting Information and associated videos).

Thresholding the foreground image and detecting connected regions of pixels allowed us to distinguish moving particles and estimate the positions of their centroids. Bed and water elevations were also deduced from the background and foreground images. Centroid positions were then tracked from frame to frame, and stacked in ``tracklets'' (i.e., parts of entire trajectories) using simple tracking rules (i.e., nearest neighbor and maximum allowed displacement). Any ambiguity arising in this tracking process was ruled out by cutting off the current tracklet and creating a new one.

In a second pass, tracklets were merged to reconstruct whole trajectories. Whenever measurements were missing (e.g., when we lost track of particles over a few frames), we used a motion model to fill information gaps. To optimize particle path reconstruction, we solved a global optimization problem using the Jonker-Volgenant algorithm \citep{Jonker1987}.
Further information is provided in the supporting information. A Matlab routine as well as a working video sample are available online ({https://goo.gl/p4GbsR}). A typical particle trajectory from start to rest is shown in a video (see heyman-ms03.avi in the Supporting Information).  The tracking algorithm was carefully validated by visual inspection and the computed average transport rates were compared to simultaneous acoustic measurements (see the supporting information). Both methods gave similar sediment transport rates, confirming the reliability of the tracking algorithm used.

\subsection{Particle Kinematics}
\begin{figure*}
\includegraphics[width=16cm]{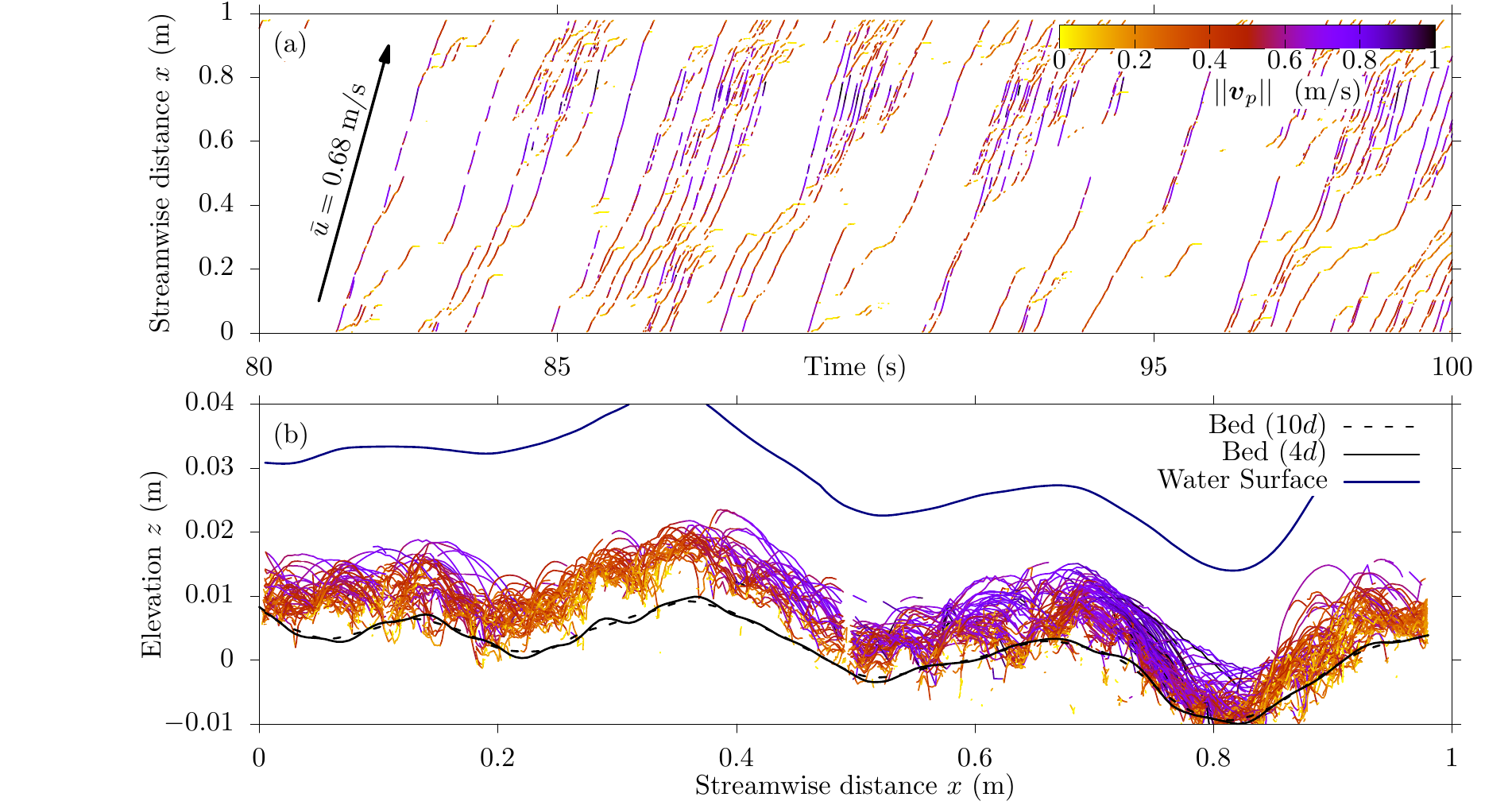}
\caption{(a) Spatio-temporal particle tracklets obtained by the tracking algorithm. The particle velocity magnitude ($\boldsymbol{v_p}$) was used as the color index. The direction of the depth-averaged flow velocity ($\bar{u}$) is indicated by the black arrow. Trajectories are limited in space by the observation window's length  (approximately 1~m along the $x$-axis). (b) Same tracklets as in plot (a), but plotted in the $(x,z)$ plane. Bed and water elevations are also plotted.  Bed elevations were smoothed out by taking the moving average over lengths $4d$ (solid line) or $10d$ (dashed line). By representing the bed surface as a smoothed curve, we ignore the local topographic details. That explains why some moving particles seemed penetrated the bed (they were above the actual bed surface, but underneath the smoothed one). }
\label{figure1}
\end{figure*}

The data resulting from image processing consisted of a set of particle positions $\boldsymbol{x}_p(t)$, where $p$ is the particle index and $t$ is the frame number, projected on a two-dimensional $({x},{z})$ plane parallel to the flume walls. An example of data is shown in Fig.~\ref{figure1} for 20~s of an experimental run.

Particle velocities ($\boldsymbol{v}_p$) and accelerations ($\boldsymbol{a}_p$) were computed using a second-order finite-difference scheme
\begin{eqnarray}
\boldsymbol{v}_p(t)&=&\frac{\boldsymbol{x}_p(t+\Delta t)-\boldsymbol{x}_p(t-\Delta t)}{2\Delta t}\\
\boldsymbol{a}_p(t)&=&\frac{\boldsymbol{x}_p(t+\Delta t)+\boldsymbol{x}_p(t-\Delta t)-2\boldsymbol{x}_p(t)}{\Delta t^2}
\end{eqnarray}
where $\Delta t=0.005$~s. The horizontal and vertical components of the state vectors are denoted as $\boldsymbol{x}_p=(x_p,z_p)$, $\boldsymbol{v}_p=(u_p,w_p)$ and $\boldsymbol{a}_p=(a_{x,p},a_{z,p})$.
Given image resolution (0.4~mm/pixel) and the frame rate (200~fps), one pixel displacement corresponded roughly to 0.08~m/s. In practice, however, particle velocity increments as small as 0.002~m/s could be recorded since the particle centroid was defined at a sub-pixel resolution as the barycenter of pixels belonging to the particle. To avoid any bias, we computed $\boldsymbol{v}_p(t)$ and $\boldsymbol{a}_p(t)$ from the tracklets obtained in the first pass of the tracking algorithm, thereby leaving aside the interpolated parts of the trajectories built during the second pass.

\subsection{Particle Activity and Flux}
Local particle activity $\gamma(x,y,t)$ (i.e., the number of moving particles per unit streambed area), was estimated from the position of moving particles. The cross-stream position $y$ of particles was not resolved in our experimental setup, thus the particle activity was expressed per unit flume width and denoted by $\gamma(x,t)$ [particles~m$^{-2}$]. Since moving particles are defined as points in space,  the local concentration in moving particles had to be computed using smoothing techniques. This was achieved by weighting particles positions with a smoothing kernel function:
\begin{eqnarray}
\gamma(x,t)=\frac{1}{B} \sum_{p=1}^{N(t)} \mathcal{K}_{\Delta}\left[x-x_p(t)\right]
\label{eq:density}
\end{eqnarray}
where $B$ was the flume width, $N(t)$ was the number of moving particles at time $t$ and $\mathcal{K}_{\Delta}$ was a smoothing kernel of bandwidth $\Delta$ [m] \citep{DiggleBook}.  Similarly, local sediment transport rate [m$^2$/s] was obtained by taking the product of particle activity and velocity:
\begin{eqnarray}
q_s(x,t)=\frac{V_p}{B} \sum_{p=1}^{N(t)} \mathcal{K}_{\Delta}\left[x-x_p(t)\right]u_p(t)
\label{eq:Kdischarge}
\end{eqnarray}
with $V_p$ the particle volume. When $\mathcal{K}_{\Delta}$ is a box filter (i.e., 0 everywhere except on a interval of length ${\Delta}$ over which it is constant), Eqs.~$\eqref{eq:density}$--$\eqref{eq:Kdischarge}$ reduce to classical volume averaging \citep{ancey02b,ancey08a,ancey14}. In the limit ${\Delta} \to 0$, the traditional equation \eqref{eq:flux-mmc} of the sediment flux as the product of a surface and a velocity is recovered \citep{heyman14thesis}. In our treatment, we chose a Gaussian kernel of bandwidth $5d$ to compute the local particle activity.

\subsection{Entrainment and Deposition}
We call ``entrainment'' the setting in motion of a bed particle. Conversely, we call ``deposition'' the passage from motion to rest of a moving particle.

In practice, an entrainment (deposition, respectively) event was detected from a particle trajectory when two conditions were met: (i) the particle velocity magnitude exceeded (fell below, respectively) a given velocity threshold $v_\text{th}$, and (ii) the distance from the particle center of mass to the estimated bed elevation did not exceed $z_\text{max}$. The first condition is usual in tracking experiments \citep{radice2006,ancey08a,campagnol2013}, and the second aims to limit overestimation of particle entrainment rates caused by broken trajectories (i.e., trajectories that were not entirely retrieved by the algorithm; see the Supporting Information for further detail).  We used a moving average value of the particle velocity over 3 frames to eliminate short periods during which a particle stayed at the same place (e.g., after a collision with the bed).
We fixed the velocity threshold at a low value ($v_\text{th}=0.01w_s$, about 5.5~mm/s) and the elevation threshold to one grain diameter above the zero bed elevation. This choice was arbitrary, but measurements of entrainment and deposition rates depended to a low degree on the velocity and elevation thresholds.

We define $P(\downarrow,\Delta t)$ as the probability that a moving particle be deposited during the short time span $\Delta t$. The particle deposition rate $D_p$ [s$^{-1}$] is then directly $D_p = {P(\downarrow,\Delta t)}/{\Delta t}$. The subscript $p$ indicates that the rate refers to a single particle. Assuming that all particles are similar, the areal deposition rate (i.e., the number of particles deposited per unit streambed area and per unit time) is simply $D=D_p \gamma$ [particles~m$^{-2}$s$^{-1}$].

Similarly,  $P(\uparrow,\Delta t)$ is the probability that a bed particle be entrained during $\Delta t$. The particle entrainment rate then reads $E_p = {P(\uparrow,\Delta t)}/{\Delta t}$ [s$^{-1}$]. Assuming a constant areal density of particles resting on the bed $\psi$, the probability that any resting particle be entrained in the infinitesimal time interval $\Delta t$, on a bed surface $S$, is $\psi S P(\uparrow,\Delta t)$. Thus, the areal entrainment rate of particles per unit bed area, denoted by $E$, is $E=\psi E_p$ [particles~m$^{-2}$s$^{-1}$].

\subsection{Hydraulic Variables}
In addition to the particle positions, we extracted the bed elevation $b(x,t)$, the local bed slope $\tan \theta(x,t)=\partial_x b(x,t)$, the water surface elevation $w(x,t)$, and its local slope $\partial_x w(x,t)$ from the images. The details of the numerical procedure to extract the flow and bed variables are given in the Supporting Information.  Note that, in contrast to the common convention in bedload transport, slopes were considered negative if the surface went downward, and positive in the opposite case.

Estimation of bed elevation and slope depended a great deal on the length scale of observation. As the bed was made of coarse particles, high fluctuations of bed slope were likely to occur when the observation scale was close to the grain diameter. Special care was paid to averaging since results were scale dependent. In practice, we selected three averaging scales $k$, corresponding to $k=4$, 10 and 40 grain diameters (i.e., 2.6, 6.7, and 26.8~cm). At each position $x$, a linear regression equation was computed based on the points located within a distance $k d/2$ from $x$. Average bed elevation and slope were then obtained from the regression coefficients. An example of the typical data obtained is shown in Fig.~\ref{figure1}b for an experimental run.

We estimated the local water depth, depth-averaged flow velocity, and bed shear stress as follows. Water depth was simply the difference between the free surface and bed elevations: $h(x,t)=w(x,t)-b(x,t)$. For stationary flows, the depth-averaged flow velocity was $\bar{u}(x,t)={Q}/({B h(x,t)})$, with $Q$  the  water inflow rate and $B$ as the channel width. Estimation of bottom shear stress was more demanding. Flow resistance resulted mainly from  friction with the glass sidewall and skin friction (related to bed roughness). The bedform's contribution to flow resistance was expected to be small compared to the two contributions above, as the wake effects induced by supercritical flows above bedforms were negligible (no flow separation, no vortices past the bedform crest).

In order to estimate wall and bed friction, we divided the  hydraulic radius $R_h$ into a ``wall'' hydraulic radius ($R_w$) and a ``bed'' hydraulic radius ($R_b$) \citep{Guo2014(1)}. The local bed shear stress was then estimated using
\begin{eqnarray}
\frac{\tau_b(x,t)}{\varrho_f} =\frac{f_b}{8}  \bar{u}(x,t)^2
\label{eq:Darcy}
\end{eqnarray}
where $f_b$ is the gravel bed's Darcy-Weisbach friction factor, which can be expressed as a function of dimensionless numbers characterizing the flow. For instance, $f_b=8\kappa^2/\ln^2(11R_b/k_s)$ in Keulegan's equation \citep{keul}, and $f_b=8 g  / (K^2 R_b^{1/3}) $ in the Manning-Strickler equation, where $\kappa$ is the von K\'arm\'an constant, $k_s$ is the equivalent roughness, and $K$ is the Strickler coefficient. Comparison of several parametrizations for $f_b$ showed that they all yielded similar estimates of bed shear stress (see the supporting information). We thus decided to use Keulegan's  equation with $k_s= 2 d$, chosen so that the friction slope matched  the bed slope on average.

\subsection{Statistical Approach}
In contrast with common practice, we did not compute sediment transport statistics for each experimental run taken separately, but we pooled the whole experimental dataset. In doing so, we envisioned our large sample (involving 5 million values) as random outcomes of the same process. In other words, statistical properties of particle paths were computed with respect to local flow and bed characteristics (i.e., local shear velocity and bed slope), but irrespectively of the overall features imposed on each experimental run (i.e, mean bed slope and water inflow rate). Pooling data was necessary since transport conditions varied significantly in time and space (Fig.~\ref{figure1}b), precluding averaging over each experimental run.

The statistical methods used in the following are standard. We focused on estimating the first and second moments (averages and variances). A subtlety arose when estimating the effects of local flow and bed slope on particle entrainment and deposition rates. Indeed, the local flow characteristics were continuous in time and space, whereas entrainment and deposition were events occurring at discrete times and places. Bayes' theorem provided a simple solution to this estimation problem. For instance, to estimate the dependence of the deposition rate on bed slope,  Bayes' theorem states that $D_p(\theta)= \left\langle{D_p}\right\rangle f_{\Theta|\downarrow}/f_{\Theta}$, where $\left\langle{D_p}\right\rangle$ was the average particle deposition rate (obtained by dividing the total number of deposition events by the total time of particle motion), $f_{\Theta|\downarrow}$ is the probability density function of bed slope associated with particle deposition and $f_{\Theta}$ was the probability density function of all bed slopes visited by moving particles (see the proof in the supporting information). The same formula applied to local areal entrainment rates, with the difference that the probability $f_{\Theta}$ should not be computed from bed slopes visited by moving particle, but from all possible bed slopes. The difference between entrainment and deposition reflected that particles could be entrained from any place along the bed surface whereas they could deposit only at the places they were visiting.

\section{Results}

\begin{table*}
\caption{Expressions and typical experimental values of the dimensionless numbers related to flow and sediment transport\tablenotemark{*}.}
\begin{tabular}[t]{rlll}
 & Name & Expression & Mean value \\
\hline
s& Density ratio & ${\varrho_p}/{\varrho_f}$ & 2.5 \\
Re& Flume Reynolds number & $4 R_h \bar{u} / \nu$& $3.5 \times 10^4$\\
Re$_*$& Particle Reynolds number & $d u_* / \nu$ & 630 \\
Re$_s$&  Reynolds  number for settling particles & $d w_s / \nu$ & 3520 \\
Fr& Froude number & $\bar{u}/\sqrt{g h \cos\theta}$ & 1.1 to 1.3 \\
$d^*$ & Dimensionless grain diameter & $d \left((s-1){g}/{\nu^2}\right)^{1/3}$ & 157 \\
--& Flume aspect ratio & $B/h$ & 0.5 to 1.5\\
--& Relative water depth & $h/d$ & 3 to 4 \\
$w_s^*$ & Dimensionless settling velocity &  $ {w_s}/{\left((s-1)g \nu\right)^{1/3}}$ & 22.5 \\
P& Rouse number & $w_s / (\kappa u_*)$ &13 \\
St& Stokes number&$Re_s s/9 $& 977 \\
$\tau^*$& Shields number& $\tau_b/( \varrho_f (s-1) g d)$ & 0.09 to 0.12 \\
$q_s^*$&  Dimensionless sediment discharge&$q_s/\sqrt{g d^3 (s-1)}$ & 2 to 7$ \times 10^{-3}$ \\
$\gamma^*$ & Dimensionless particle activity & $\gamma  d^2$ &4 to 12 $\cdot 10^{-3}$ \\
$E^*$& Dimensionless areal entrainment rate & $E d^3/ w_s $& $1.04 \times 10^{-4}$\\
$D^*$& Dimensionless areal deposition rate & $D d^3/ w_s$ & $1.04  \times 10^{-4}$\\
$D_p^*$& Dimensionless particle deposition rate & $D_p d/ w_s $& 0.0148\\
\hline
\end{tabular}
\tablenotetext{*}{$\varrho_f$ is the fluid density, $\varrho_p$ the particle density, $R_h$ the hydraulic radius, $\bar{u}$ the depth-average water velocity, $\nu$ the fluid kinematic viscosity, $g$ gravitational acceleration, $h$ the water depth, $\theta$ the bed slope, $d$ particle mean diameter, $u_*$ the shear velocity, $B$ the channel width, $\kappa$ von K\'arm\'an's constant, $w_s=0.55$ m/s the settling velocity determined from Eq.~(37) in \citet{Brown2003}, $\tau_b$  the bed shear stress, $q_s$  the sediment transport rate, $\gamma$ the particle activity (i.e., the concentration in moving particles), $E$ the areal entrainment rate; $D$ the areal deposition rate, and $D_p$ the particle deposition rate.}
\label{tab:adim}
\end{table*}

\begin{figure*}
\includegraphics[width=16cm]{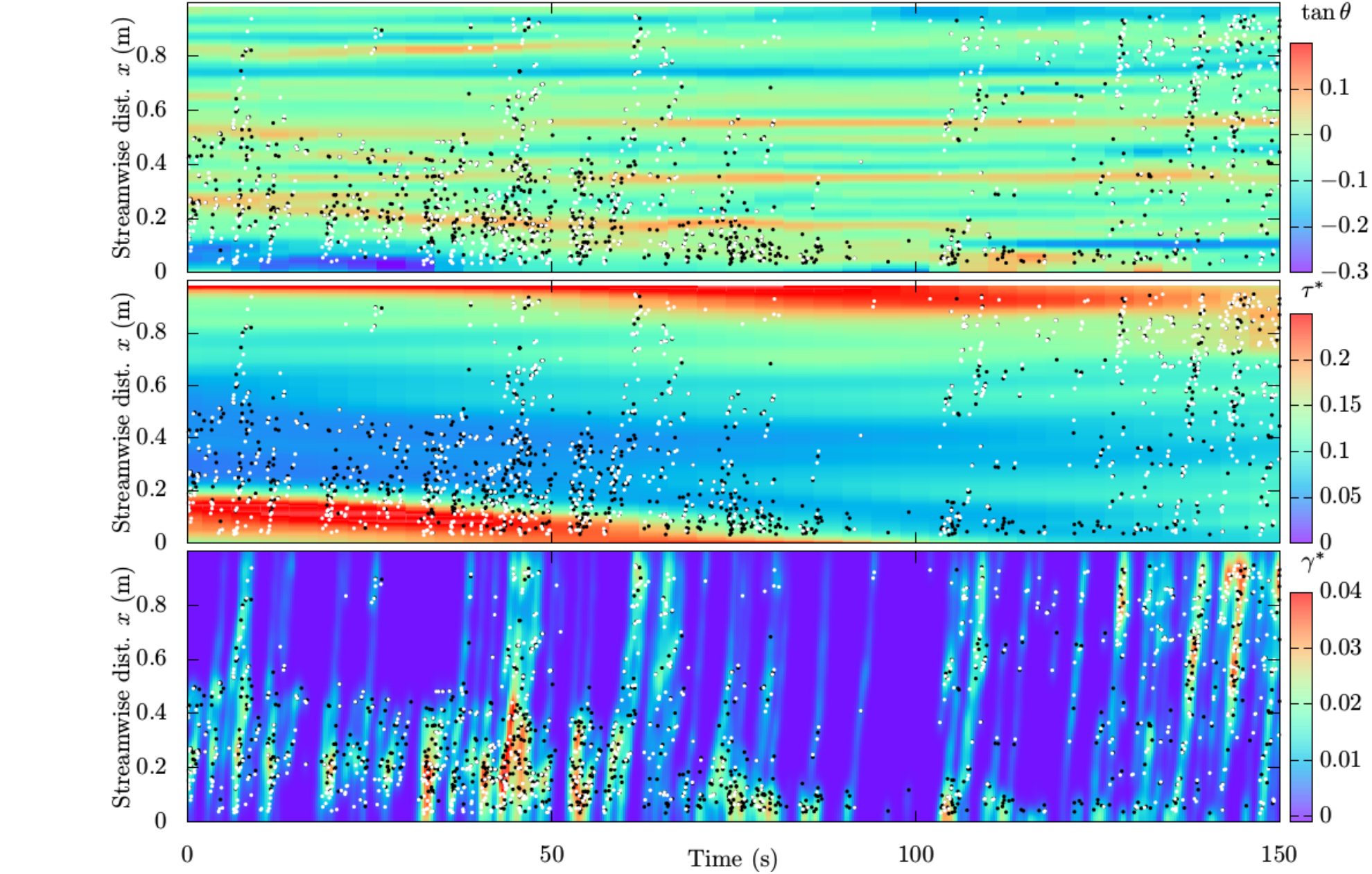}
\caption{Spatio-temporal variations in bed slope $\tan \theta$ (top), Shields number (middle) and dimensionless particle activity (bottom). Flow direction is from bottom to top. The local averages  of bed slope and Shields number were computed over a length scale $k=10d$, with $d$ the mean grain diameter. The bandwidth of the Gaussian kernel used to compute $\gamma^*$ is $\Delta=5d$ (see Eq.~\eqref{eq:density}). Particle entrainment and particle deposition events are indicated by white and black dots, respectively.}
\label{figure2}
\end{figure*}

\subsection{Flow Conditions}
Table~\ref{tab:adim} summarizes the dimensionless numbers related to sediment transport and their typical ranges of variation in our experiments (details about each experimental runs are given in the Supporting Information).
The high Reynolds numbers observed in our experiments suggest the occurrence of a fully turbulent flow with a rough hydrodynamic bed boundary.  Froude numbers above unity show that flow was supercritical. The water depth was only three or four time greater than the grain diameter. These shallow water conditions are frequently encountered in gravel-bed and mountain streams.

The Rouse number P is used in sedimentation studies to evaluate the propensity of particles to be carried in suspension by  turbulence. It relates the particle settling velocity $w_s$  to an estimate of the upward fluctuating velocity $\kappa u_*$, with $\kappa$ the von K\'arm\'an constant. The observed mean values of P were large ($\mathrm{P}\approx 13$) indicating that all particles were transported as bedload. The Stokes number St is introduced in two-phase flows to evaluate the strength of the coupling between the water and particle phases \citep{batchelor89}. It may be interpreted as the ratio between the particle relaxation time and flow timescale. When $\mathrm{St}\gg1$, the fluid has no time to adjust its velocity to variations in particle velocity and, conversely, the particle is not affected by rapid variations in the fluid velocity (but, naturally, it continues to be affected by the slow variations). In practice, this means that the fluid and particle move in a quasi-autonomous way. On a macroscopic scale, such suspensions retain a genuinely two-phase character and the equations of motion take the form of two interrelated equations (one for each phase). When $\mathrm{St}\ll1$, the particle has time to adjust its velocity to any change in the fluid velocity field. One sometimes says that the particle is the slave of the fluid phase. On a macroscopic scale, this means that the suspension behaves as a one-phase medium. The Stokes number is also used to evaluate the effect of fluid viscosity on particle collision in water: large St are generally associated with low viscous dissipation \citep{schmeeckle01,joseph01}. In our experiments, the large values of St found ($\sim 1000$) indicate that (i) collisions were elastic with little viscous damping and (ii) bedload transport behaved like a two-phase system.

Sediment transport rates were fairly low in our experimental campaign, as shown by the small dimensionless activity and sediment transport rate. On average, only 0.4\% to 1.2\% of the bed was covered with moving particles. This contrasts with what \citet{lajeunesse10} observed: they explored regimes in which at least 5\% of the bed surface was mobile (in their experiments, $-\tan \theta \approx 0-11$~\%, $d\approx1-5$ mm). In the present study, a bedload sheet layer never developed and particles moved sporadically by little jumps.

Figure~\ref{figure1} shows that flow, bed, and sediment transport varied locally, even though the water discharge and sediment feed rate were kept constant. Small bedforms developed naturally,  modifying local bed topography, flow, and sediment transport. The typical length of these structures ranged from 5$h$ to 10$h$, i.e. from 10 to 20~cm, while their amplitude was equal or smaller than the water depth $h$. Their celerity depended on the sediment transport rate, but were always low compared to particle and flow velocities. In Fig.~\ref{figure2}, we plotted the spatio-temporal variation of bed slope, bed shear stress and local particle activity in one experiment where bedforms were seen moving upstream. We emphasize that the presence of these bedforms induced local modifications in flow and sediment transport, providing grounds for the statistical approach taken in this paper.

\subsection{Kinematics}
Particles mainly traveled downstream by saltating or by rolling on the granular bed. Their velocities exhibited large fluctuations and, thus, they  spread along the bed. In the following, we successively present particle accelerations, particle velocities and particle spreading (i.e., diffusivity).

\subsubsection{Particle Acceleration}
\begin{figure*}
\includegraphics[width=16cm]{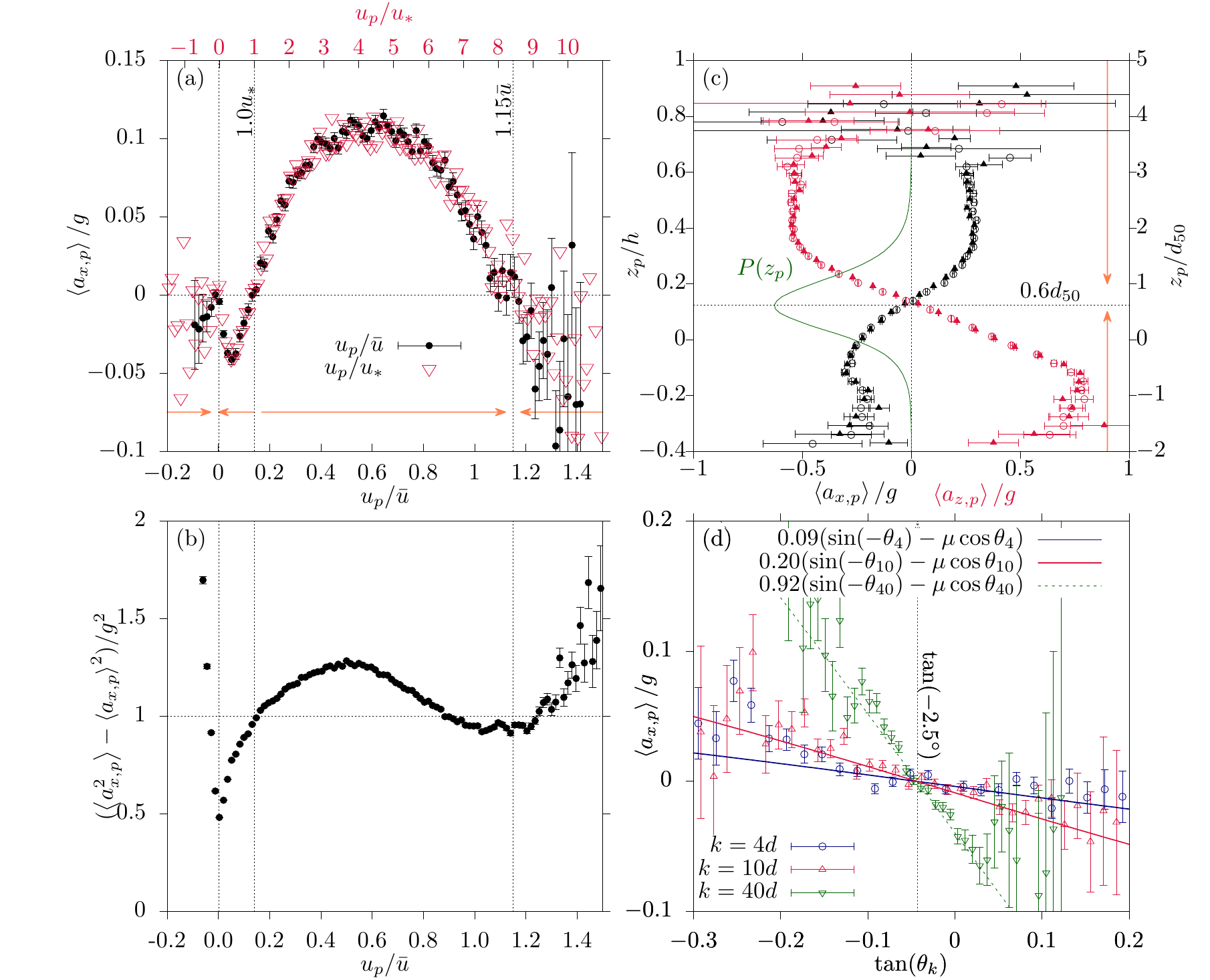}
\caption{(a) First and (b) second moments of the streamwise component of particle acceleration ($a_{x,p}$) as a function of particle velocity ($u_p$) scaled by the depth averaged flow velocity ($\bar{u}$) or the shear velocity ($u_*$). The gravitational acceleration is denoted by $g$. (c) Average streamwise component and vertical component of particle acceleration ($a_{x,p}$ and $a_{z,p}$ respectively) as a function of particle elevation $z_p$ scaled either by water depth ($z_p/h$, circles) or by particle mean diameter ($z_p/d$, triangles). Arrows indicate the direction of the average momentum change depending on particle velocity. (d) Streamwise component of particle acceleration as a function of bed slope ($\theta$). Bed slope was computed by taking the moving average of local slopes over various length scales $k$: $4d$, $10d$ and $40d$.}
\label{figure3}
\end{figure*}

The study of particle acceleration provides some insight into the nature and magnitude of forces driving bedload particles. Figure~\ref{figure3}a shows that the ensemble-averaged streamwise component of particle acceleration  $\left\langle{a_{x,p}}\right\rangle$  varied nonlinearly with $u_p$. Scaling particle velocity with $u_*$ or $\bar{u}$ gave similar trends. $\left\langle{a_{x,p}}\right\rangle$ changed sign depending on the instantaneous particle velocity. Deceleration dominated for $0<u_p<u_*$ and for $u_p>1.15\bar{u}$ (total force resisting motion), whereas particles were mostly accelerating for $u_*<u_p<1.15\bar{u}$ (total force promoting motion). When $u_p<0$, the wide scatter of points makes the analysis difficult, but it is likely that $\left\langle{a_{x,p}}\right\rangle>0$ since particles that temporarily moved backward (after a collision for instance) quickly followed the flow direction again.

Streamwise particle accelerations showed large variations around the mean (Fig.~\ref{figure3}b). Indeed, the standard deviation of streamwise acceleration was always close to $g$, whereas the mean never exceeded $0.1g$ (Fig.~\ref{figure3}a and b). Variations were strongest at relatively low ($u_p<0$) or high particle velocities ($u_p>1.15\bar{u}$). We distinguished two local minima, at $u_p=0$ and $u_p\approx1.15\bar{u}$, as well as a local maximum near $\bar{u}/2$.

We plotted the dependence of streamwise and vertical accelerations on particle elevation in Fig.~\ref{figure3}c. The change of sign in vertical particle acceleration at about $z_p=0.6d$ suggests that most particles moving below $0.6d$ experienced upward acceleration, whereas above this elevation, they were subject to negative acceleration of the order of $-g/2$. As previously noted with the distribution of particle elevations,  particles were  mainly transported at an elevation of $z_p\approx 0.6d$. Interestingly, both vertical and streamwise particle accelerations changed sign at the same elevation and always remained of opposite sign.

Streamwise particle acceleration depended almost linearly on local bed slope (Fig.~\ref{figure3}d), the largest accelerations being observed on the steepest slopes. Acceleration became negative when $\tan \theta$ was larger than $-\tan  2.5^\circ  = -0.043$, e.g., for shallow slopes and adverse bed slopes.  Note that the scale at which the bed slope was computed influenced the relationship: the larger the length scale, the stronger the dependence.

\subsubsection{Particle Velocity}
\begin{figure*}
\includegraphics[width=16cm]{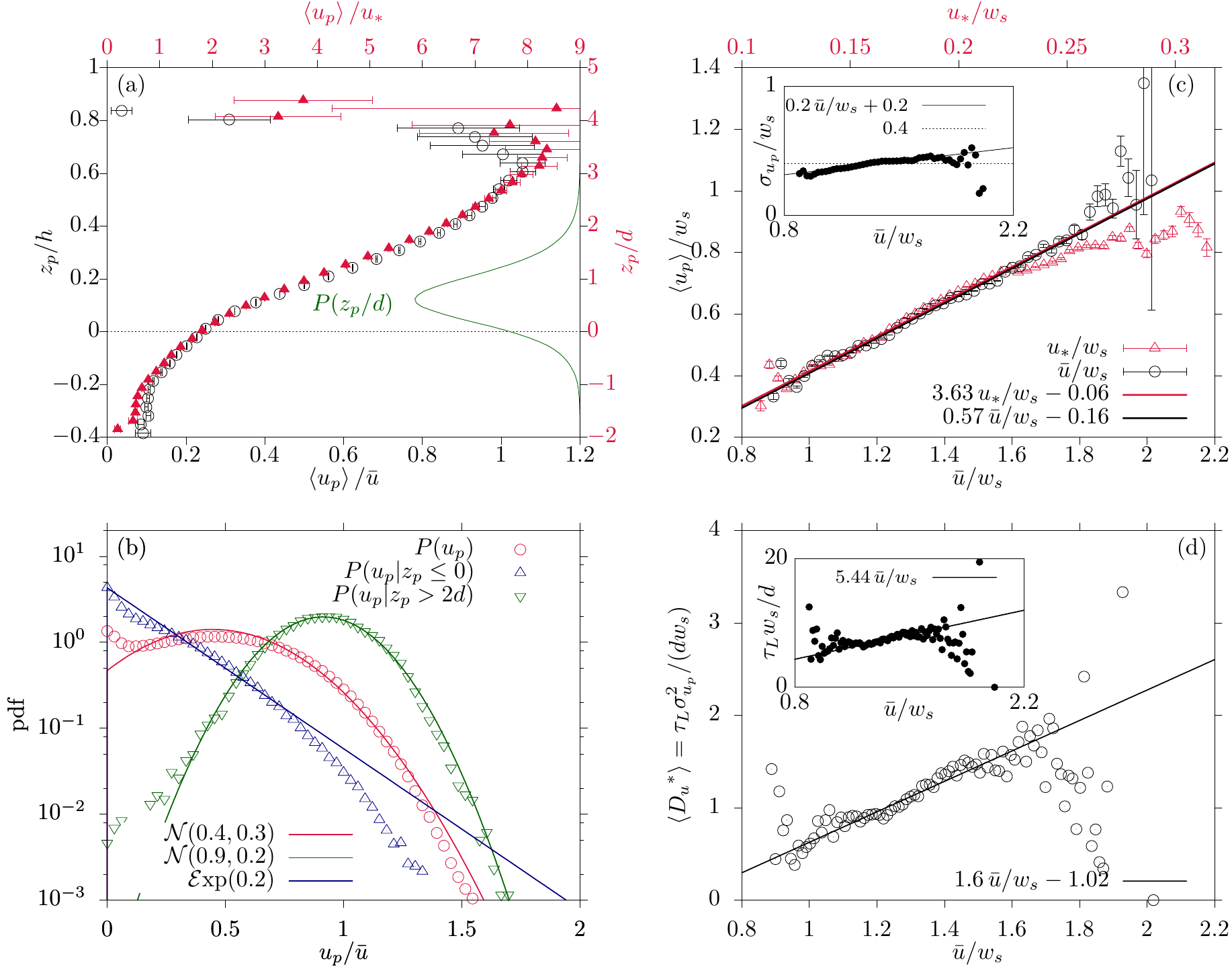}
\caption{(a) Average particle velocity ($u_p$) as a function of particle elevation ($z_p$). Red triangles:  $\left\langle u_p\right\rangle/u_* $ as a function of $z_p/d$, where $u_*$ is the shear velocity and $d$ the mean grain diameter; Black circles: $\left\langle{u_p}\right\rangle/\bar{u}$ as a function of $z_p/h$, where $\bar{u}$ is the depth average flow velocity and $h$ the flow depth; Green line: distribution of moving particle elevations as a function of $z_p/d$. (b) Distribution of instantaneous particle velocity and theoretical fits. $\mathcal{N}(\mu,\sigma)$ is the Gaussian distribution of mean $\mu$ and standard deviation $\sigma$ and $\mathcal{E}$xp($\lambda$) is the exponential distribution with mean $\lambda$. (c) Average particle velocity as a function of $u_*$ (blue triangles) and $\bar{u}$ (black circles). All velocities are scaled with the particle settling velocity ($w_s$). Inset: Standard deviation of particle velocity as a function of  $\bar{u}$. (d) Dimensionless particle diffusivity ($D^*_u$) as a function of $\bar{u}$. Insets: Lagrangian time scale ($\tau_L $) as a function of $\bar{u}$.}
\label{figure4}
\end{figure*}

Figure~\ref{figure4}a shows that the ensemble-averaged streamwise component of particle velocity $\left\langle{u_p}\right\rangle$ depended on its elevation above the gravel bed ($z_p$). Low velocities (of the order of $u_*$) were found close to the bed while larger velocities (of the order of $\bar{u}$) were observed at higher elevations above the bed.

As shown in Fig.~\ref{figure4}a, moving grains were observed at elevations as low as $z_p=-2d$. In fact, $z_p$ was only a rough estimate of particle elevation over the granular bed, since the position of the latter was not known exactly. The distribution of moving particle elevations $P(z_p)$ (see Fig.~\ref{figure4}a) suggested that most of the time, particle hop amplitude was limited to $2$ particle diameters above the zero bed level. Between these two bounds,  average streamwise particle velocity varied approximately by a factor of 6.

Instantaneous streamwise velocities also showed large variations, fairly well described by a truncated Gaussian distribution (Fig.~\ref{figure4}b).  Interestingly, the velocity of particles moving at low elevations ($z_p<0$) was exponentially distributed, whereas the velocity of particles moving at high elevation ($z_p>2d$) was clearly Gaussian.

A linear relationship was observed between particle velocity and both $u_*$ and $\bar{u}$ (Fig.~\ref{figure4}c). Here, all velocities were made dimensionless by using the settling velocity $w_s$. Note that, at high velocities, the linear fit was better when $\left\langle{u_p}\right\rangle$ was related to the depth-averaged flow velocity $\bar{u}$.

\subsubsection{Particle Diffusivity}\label{sec:diff}
Particle diffusivity can be obtained by two methods: (i) by computing the mean squared particle displacement (which grows asymptotically as $2D_u t$ in the case of normal diffusion) or (ii) by estimating separately the Lagrangian integral timescale of the particle velocity time series $\tau_L$ (i.e., the integral of the particle velocity autocorrelation function $\rho_{u_p}(t)$) and the variance of particle velocity $\sigma_{u_p}^2$. It can be shown that $D_u=\tau_L \sigma_{u_p}^2$ in the case of normal diffusion \citep{Furbish2012(4)}.

The first method does not apply to nonuniform or unsteady flows since it is based on the asymptotic scaling of displacements, and thus ignores local variations in the transport process. In contrast, an estimation of $\tau_L$ for non-uniform flows is possible if we assume that the autocorrelation function of particle velocity is closely described by an exponential function: $\rho_{u_p}(t) \approx \exp(-t/ \tau_L)$ \citep{martin12}. Consequently, $\tau_L$ can be computed based on the very first lags of the particle velocity autocorrelation function. The estimation of $\sigma_{u_p}$ from the trajectories is straightforward.
In our case, we found that both $\tau_L$ and $\sigma_{u_p}$ increased slightly with depth-averaged flow velocity  (inset of Fig.~\ref{figure4}c and d).

We introduce the dimensionless particle diffusivity $D_u^*=D_u/ (dw_s)$ and plot its variations against the depth-averaged flow velocity in Fig.~\ref{figure4}d. $D_u^*$ increased almost linearly with $\bar{u}/w_s$, from 0.5 at low shear velocities to 2 at higher shear velocities. The trend at high flow velocities reflected a decay of particle diffusivity, but the scatter of data made it difficult to draw sound conclusions.

\subsection{Mass Exchange}
\begin{figure*}
\includegraphics[width=16cm]{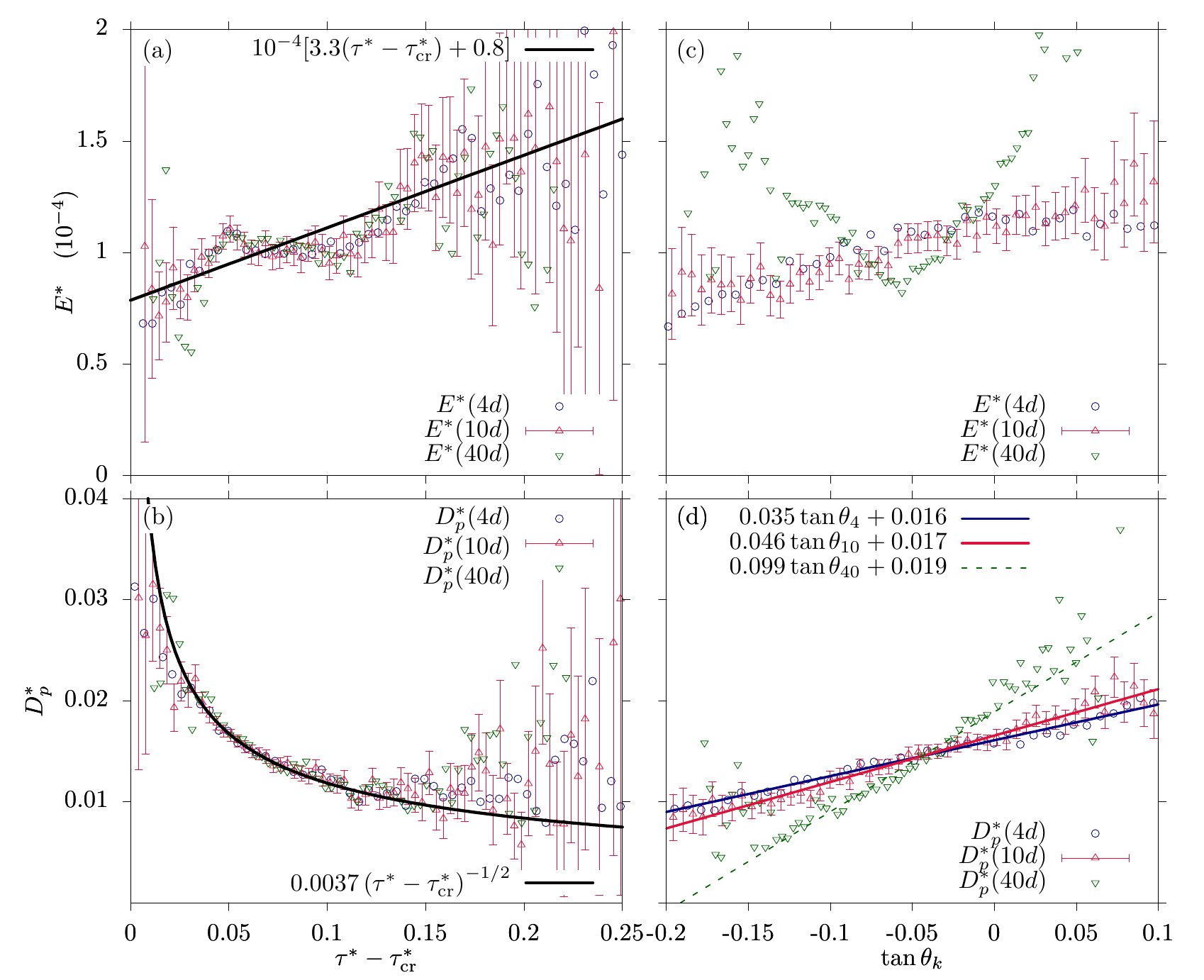}
\caption{ (a) Dimensionless areal entrainment rate ($E^*$) and (b) particle deposition rate ($D_p^*$) as a function of the excess Shields number $\tau^*-\tau_\text{cr}^*$, with $\tau_\text{cr}^*= \tau_{0}^* \sin\left( \theta+\alpha \right)/\sin \alpha$, $\alpha=47^\circ$ and $\tau_{0}^*=0.035$. (c) Areal entrainment rate and (d) particle deposition rate as a function of the local bed slope $\theta_k$, where $k$ is the averaging scale. $d$ is the mean particle diameter.}
\label{figure6}
\end{figure*}

At low sediment transport rates, particle motion was intermittent and few particles moved: between long periods of rest, a particle was occasionally picked up, transported, and deposited  farther downstream. The irregular particle shape precluded creeping motion observed with granular packings made of glass beads \citep{houssais15}, and so individual particle entrainment and deposition were the dominant processes  of mass exchange.

We first take a look at the dependence of entrainment and deposition rates on the local \emph{excess} Shields number, defined as $\tau^*-\tau_\text{cr}^*$, where $\tau_\text{cr}^*$ is given by \citet{FLvB}: $\tau_\text{cr}^*= \tau_{0}^*\sin\left( \theta+\alpha \right)/\sin \alpha$, with $\alpha \approx 47^\circ$ and $\tau_{0}^*=0.035$, a reference angle of repose and the critical (or reference) Shields number at zero slope respectively.

The areal particle entrainment rate did not increase significantly with $\tau^*-\tau_\text{cr}^*$; for excess shear numbers in the 0.05--0.1 range, $E^*$ showed even a slight decrease (Fig.~\ref{figure6}a). By contrast, Fig.~\ref{figure6}b shows a strong inverse correlation between the  particle deposition rate and the excess Shields number: $D^*$ varied between about 0.01 at high excess Shields numbers to 0.03 at low excess Shields numbers. Note also that the length scale over which bed slope was measured did not significantly affect the results.

No clear relationship was observed between $E^*$ and the local bed slope: at small length scales ($k\leq 10d$), particle entrainment rates were higher on shallower bed slopes, but at large length scales ($k\geq 40d$), this trend no longer held (Fig.~\ref{figure6}c). On the opposite, particle deposition rates followed clearer trends. They increased linearly with decreasing bed slope at all length scales: the steepest the slope, the lower the deposition rate (Fig.~\ref{figure6}d). Moreover, the larger the length scale, the heavier the dependence of the particle deposition rate  on $\theta$ .

\begin{figure}
\includegraphics[width=8cm]{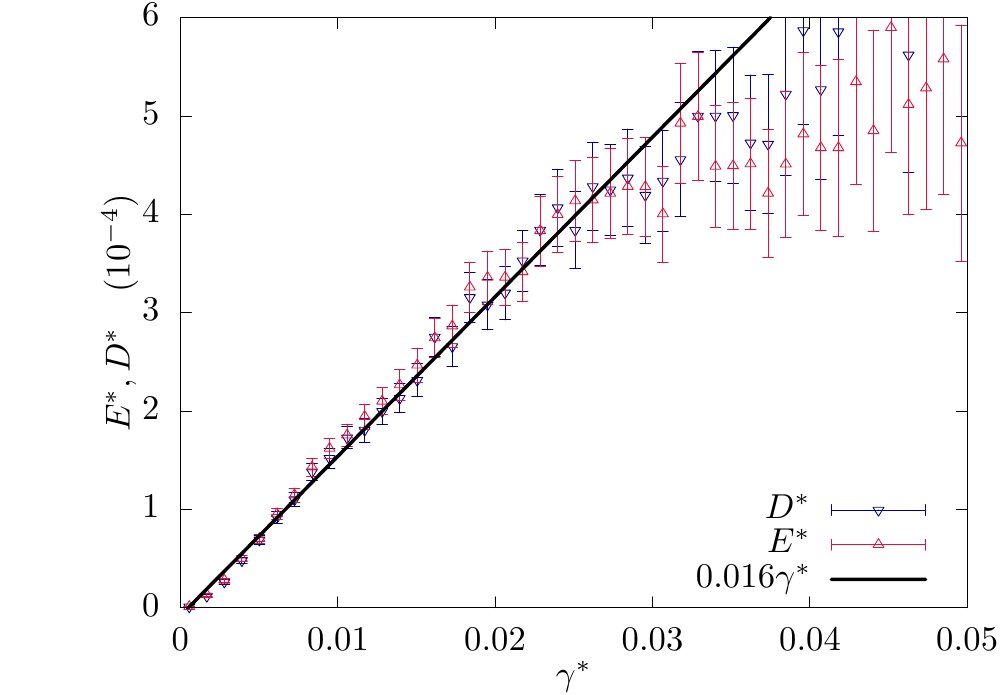}
\caption{Dimensionless areal entrainment ($E^*$) and areal deposition rates ($D^*$) as a function of local dimensionless particle activity ($\gamma^*$). The latter is estimated from Eq.~\eqref{eq:density} taken with a Gaussian kernel of bandwidth $5d$, where $d$ is the mean grain diameter.}
\label{figure8}
\end{figure}

Finally, the dependence of $D$ and $E$ on local particle activity is shown in Fig.~\ref{figure8}.  Local particle activity $\gamma(x,t)$ was computed from Eq.~$\eqref{eq:density}$ with $\Delta=5d$  (see Fig.~\ref{figure2}).
It came as no surprise that the areal deposition rate depended on local particle activity (i.e., the more grains in motion, the more grains to be deposited). More interestingly,  the areal entrainment rate showed almost the same dependence  on particle activity.  A regression fitted to values of $\gamma^*<0.03$ gave a slope of approximately 0.016 for both areal deposition and entrainment rates.

\section{Discussion}
For the flow conditions and sediment characteristics encountered in our experiments, the critical Shields number for incipient motion ($\tau_\text{cr}^*$) usually fells in the $0.02-0.08$ range \citep{buffington1997}. This suggests that the transport stage $T=\tau^*/\tau_\text{cr}^*$ (i.e., the ratio of the Shields number to the critical Shields number) ranged from 1 to 6 in our experiments. More specifically, we observed that $T$ was very close to, if not below, 1. When sediment supply was stopped at the end of an experiment, sediment transport in the flume also terminated rapidly, showing that the flow alone was often not sufficiently vigorous to entrain particles, at least over the experimental timescales considered. Conversely, when sediment was supplied at the flume inlet, sediment transport started spontaneously and was maintained without net bed aggradation or degradation. In other words, the initiation threshold appeared to be significantly higher than the cessation threshold in all our experiments. These particular conditions were also reported in previous experimental studies \citep{francis73,drake88a,ancey02b}, field investigations \citep{Reid1985}, and more recently in numerical models for particles at high Stokes number \citep{Clark2015}.

\subsection{Particle Acceleration}
The total force applied on a particle was closely related to its instantaneous acceleration given by Eq.~\eqref{eq:motion2}.  All parameterizations available in the literature failed to describe the nonlinear particle dynamics we observed in our experiments (Fig.~\ref{figure3}).
In the following, we provide further elements with which to understand the specific behavior of particle acceleration in our experiments.

First, Fig.~\ref{figure3}a shows that the average streamwise acceleration passed from positive to negative values at $u_p = 0$ and $u_p \approx \bar{u}$. This suggests that two equilibrium states or ``attractors'' were possible for the streamwise component of particle velocity.  Particles came to a halt after a few displacements so that the zero velocity was obviously a natural attractor. The peak observed at $u_p=0$ in the particle velocity density function (see Fig.~\ref{figure4}b) conveys the same information. The existence of a second attractor at $u_p \approx \bar{u}$ is best seen in Fig.~\ref{figure1}a, where numerous particle paths are almost parallel to the mean flow velocity in the time-space plane. Apart from these two velocity attractors, the state $u_p \approx u_*$ appeared as a tipping point: a moving particle might, with equal probability, accelerate or decelerate, suggesting that $u_p \approx u_*$ was an unstable equilibrium state.

Interestingly, \citet{Quartier2000} showed that, for certain slopes, a cylinder rolling down a rough plane may either come to rest or reach a constant velocity depending on its initial kinetic energy, and presented an acceleration diagram very similar to the one shown in Fig.~\ref{figure3}(a). In other words,  for a grain to stay in motion it must always rise to the top of the grains beneath it. This phenomena is well documented in the granular media literature and leads to an hysteresis in the initiation and cessation of motion.  Bed troughs trap moving particles without sufficient kinetic energy \citep{riguidel94b,ancey96b,dippel97c}. We believe that a similar trapping phenomenon occurred in our experiments, which may explain the particular nonlinear shape of the longitudinal particle acceleration function.

The simultaneous change of sign in  the streamwise and vertical components of particle acceleration at an elevation $z_p \approx 0.6d$ (see Fig.~\ref{figure3}b) also suggests that part of the streamwise momentum is converted into vertical momentum, probably as a result of particle impacts on the bed. As shown by \citet{gondret99}, \citet{schmeeckle01} and \citet{joseph01}, when Stokes numbers exceed 1000, viscous forces dissipate only part of the kinetic energy and therefore, most of the momentum is restituted to the colliding particle (part of the momentum may also be transmitted to the bed particles by elastic waves). Depending on the contact angle, streamwise momentum is transformed into vertical momentum upon impact, allowing the particle to take off and to hop.

As Fig.~\ref{figure3}d shows a linear relationship between particle acceleration and bed slope, one can wonder whether there is a simple mechanical explanation for this linearity. We  assume that moving particles undergo four different forces: gravitational and buoyant forces $m' g \sin\theta$ (with $m'=(\varrho_p-\varrho_f)V_p$ the buoyant weight), Coulomb-like frictional force $\mu m'  g \cos\theta$ (with $\mu$ a Coulomb friction coefficient), the collisional force $F_c$, and drag force $F_d$. To simplify the analysis, we assume that the contact forces (the frictional and collisional forces) are bulk forces that arise from energy lost in collisions and rubbing \citep{jaeger90,ancey96b,ancey03a}, and thus they do not act only when the particle is in contact with the bed. The momentum balance equation for one particle takes the form in the streamwise direction
\begin{linenomath*}\begin{equation}
m_p  a_{x,p}    =m'g(\sin(-\theta) -\mu \cos \theta) - F_c+F_d
\end{equation}\end{linenomath*}
The last two contributions $F_c$ and  $F_d$ do not depend on slope explicitly. Taking the ensemble average and dividing by $m_p g$, we end up with
\begin{linenomath*}\begin{equation}
\frac{\left\langle{a_{x,p}}\right\rangle}{g}=\alpha(\sin(-\theta) -\mu \cos \theta) + \langle f  \rangle
\label{eq:friction}
\end{equation}\end{linenomath*}
so as to make the dependence on $\theta$ more apparent (with $\alpha=m'/m_p$ and $f=(  F_d-F_d)/(m_p g)$). Equation \eqref{eq:friction} fits the experimental data of Fig.~\ref{figure3}d if we take $\mu=\tan 2.5^\circ$, $\langle f  \rangle=0$,  and $\alpha$ ranging from 0.09 to 0.92  for bed slopes averaged over length scales between $4d$ and $40d$. How realistic is the model above? Three remarks can be made. First, the Coulomb coefficient is off by a factor of 10. This may indicate that as the particle saltates, the moments during which the particle is in contact with the bed and experiences friction, are infrequent, and thus frictional dissipation is low. Second, Eq. \eqref{eq:friction} exhibits scale dependence: it performs better for large length scales, for which $\alpha$ comes  closer to the expected value $\hat\alpha=m'/m_p\sim0.63$.  The role played by the length scale may reflect the fact that a saltating particle ``feels'' the effect of local bed slope after several contacts with the bed.  In this respect, the effective bed slope should be computed over sufficiently large length scales to be relevant for particle motion. Third, taking $\langle f  \rangle=0$ is equivalent to assuming that the drag force exerted by the water stream exactly matches the collisional force, but physically, we see no special reason for this matching. In short, a simple model like  Eq. \eqref{eq:friction} does not offer a sufficient parametrization of the forces acting on saltating particles.

\subsection{Particle Velocity}
In agreement with earlier results \citep{francis73,abbott77,vanrijn84a,nino94a,ancey02b, lajeunesse10,martin12}, we found that the average particle velocity increased almost linearly with the shear (or depth-averaged flow velocity, respectively). A linear fit to the data gives $\left\langle{u_p}\right\rangle\approx A u_* - 0.06 w_s$ with $A=3.6$ (or $\left\langle{u_p}\right\rangle \approx 0.57 \bar{u}- 0.16 w_s$, respectively). This is in close agreement with the values reported by \citet{lajeunesse10} and \citet{nino94a}, who found $A$ in the 4.4--5.6 range.
Authors, such as \citet{abbott77} and \citet{martin12} reported higher $A$ values, but their experiments were run over fixed beds.

At high flow velocities, the depth-averaged velocity $\bar{u}$ seemed a better predictor of  $\left\langle{u_p}\right\rangle$ than the shear velocity $u_*$. High particle velocities corresponded to large hop amplitudes, and so far from the bed, flow velocity was better approximated by $\bar{u}$ than $u_*$.  In addition, as the flow was shallow ($h/d\approx3.5$), the depth-averaged  velocity provided the right order of magnitude of the velocity field seen by the particles, this situation is thus quite different from deep flows for which shear velocity is representative of the water velocity in the neighborhood of moving particles.

Our results also show that, similarly to \citet{martin12} and \citet{ancey14}, the particle velocity distribution was almost Gaussian. It is worth noting  that  these experiments were run at particle Reynolds numbers $\mathrm{Re}_*$ larger than 500, whereas studies reporting an exponential distribution of particle velocity involved low particle Reynolds numbers: $\mathrm{Re}_* <100$  in \citet{lajeunesse10} and  $\mathrm{Re}_* <20$ in \citet{furbish12a}. The particle Reynolds number thus seems to be the key factor in partitioning the velocity regimes.

Yet this is not the only factor. A closer look at the results shows that particles velocities tended to be exponentially distributed near the bed while, far from the bed, their distribution was Gaussian (Fig.~\ref{figure4}b).  The transport mode---low rolling/sliding or  saltation---also controlled the shape of the particle velocity distribution.

\subsection{Diffusivity}
In this paper, we used a method---hereafter referred to as AC--- to compute particle diffusivity from the autocorrelation time and standard deviation of particle velocities. We found that the dimensionless particle diffusivity $D_u^*$ varied almost linearly with the depth-averaged flow velocity ($D_u^*=1.6 \bar{u}/w_s -1.2$), yielding $D_u^*$ values in the 0--2 range.
\citet{furbish12c} used a similar approach and obtained $D_u^* \approx 0.7-1.9$. By contrast, using the linearity of the mean squared displacement (a method referred to as MSD), \citet{heyman14} found larger values for diffusivity ($D_u^*= 2.8-5$) as did \citet{Kavvas2015} ($D_u^*\approx 3.6$).

The mismatch between MSD and AC estimates result from different working assumptions: MSD considers particle motion on long times, whereas AC is computed on short times. In numerous systems driven by fluctuations, particle motion is ballistic at short times (typically, for times shorter than the flight time between two collisions), resulting in super-diffusion  \citep{pusey11}, and this behavior is also observed for bedload transport \citep{martin12,heyman14thesis,Bialik2015}. As a result, apparent diffusivity is smaller at short timescales.

The variety of processes and timescales involved in particle spreading may also explain the differences between MSD and AC estimates \citep{Bialik2015}. Repeated impacts of particles with the bed are the primary source of velocity fluctuations. The timescale of these fluctuations depends on the impact frequency, and thus can be long for saltating particles. Turbulent drag fluctuations and variations in contact forces also modify particle path. In addition to reducing diffusivity, these processes occur on shorter timescales than particle impacts, and this is another cause of the disagreement between MSD and AC.

\subsection{Mass Exchange}

The particle deposition rate was closely related to the local excess Shields stress and bed slope: the highest deposition rates were found at low Shields number and adverse bed slopes (Fig.~\ref{figure4}b). This contrasts with previous experimental findings suggesting a globally constant deposition rate under steady uniform flow conditions \citep{ancey08a,lajeunesse10}. A nonlinear fit to the data suggests an inverse dependence of the deposition rate upon shear velocity ($D_p^*\propto u_* ^{-1}$, or equivalently $D_p^*\propto 1/\sqrt{\tau^*}$).

The inverse dependence of particle deposition rate on shear velocity is also supported by the peculiar shape of the streamwise acceleration. Figure~\ref{figure3}a shows that particles with velocities below $u_*$ were mainly decelerating, an effect that we interpreted as particle trapping by bed asperities. Thus, the lower $u_*$, the more likely a moving particle is to be trapped  and thus, the higher the deposition rate.

As expected, areal deposition rates were well correlated with local particle activity (by definition $D^*=D_p^*\gamma^*$). A linear fit gave $D_p^*\approx 0.016$, a value very close to the average particle deposition rate computed independently from  individual particle trajectories ($\left\langle{D_p^*}\right\rangle\approx 0.015$).

By contrast, no strong correlation between areal particle entrainment rate and flow strength was found: $E^*$ increased with $\tau^*-\tau_\text{cr}^*$ at a rate of $3.3\times 10^{-4}$, a value much lower than the rate found by \citet{lajeunesse10} for mild sloping beds (they found a regression coefficient of 0.43).  Interestingly, in their experiments over steep slopes, \citet{ancey08a} did not report any correlation between entrainment rates and flow strength, suggesting that either this behavior was specific to transport of coarse grains over steep slope or resulted from the narrowness of the flume.

Surprisingly, $E^*$ did not vanish when the Shields number came close to or below the estimated threshold of incipient motion. This suggests that mechanisms  other than flow drag  facilitated particle entrainment, which was confirmed by the clear linear correlation between areal entrainment rates and particle activity (Fig.~\ref{figure8}).  The correlation suggests that particle entrainment was enhanced by the passage of other moving particles.

In the supporting information, we provide a video showing the setting in motion of a bed particle due to the impact of a moving grain. As discussed previously in Sec. 5.1, when particles are characterized by high Stokes numbers, the viscous forces weakly damp momentum transfer when a moving particle impacts the bed. The amount of energy transferred to a resting particle upon impact may be sufficient to dislodge it. This energy transfer is even more effective for flows close to the threshold of incipient motion: a small amount of energy suffices to set the resting grain in motion.

To account for the feedback loop due to moving particles in the entrainment process, \citet{ancey08a} proposed breaking down the areal entrainment rate into a flow contribution $e_0$, depending on bottom shear stress, and a particle contribution $e_1$, depending on particle activity (termed the \emph{collective entrainment} rate in their original paper):
\begin{linenomath*}\begin{equation}
E^*=e_0+e_1 \gamma^*
\label{eq:eroco}
\end{equation}\end{linenomath*}
Figure~\ref{figure8} suggests that $e_1 \approx 0.016$ and that $e_0\approx 0$, confirming that particle entrainment was essentially triggered by the passage of  moving particles in our experiments.

Equation~\eqref{eq:eroco} shows that the particle activity equation can be cast in the form
\begin{linenomath*}\begin{equation}
\frac{\partial \gamma^*}{\partial t}+ \frac{1}{V_s}\frac{\partial q_s^*}{\partial x} = -\varepsilon \gamma^* + e_0
\label{eq:gamma2}
\end{equation}\end{linenomath*}
where $\varepsilon=D_p^*-e_1$. The values of $D_p^*$ and $e_1$ were very close ($D_p^*\approx e_1\approx 0.016$), thus $\varepsilon \approx 0$ and the source term in Eq.~$\eqref{eq:gamma2}$ canceled out. Particle activity at equilibrium was thus solely dictated by the imposed upstream boundary condition.  In other words,  regardless of the transport capacity of the flow, the particle flux matched the sediment supply rate, and the bed remained near equilibrium. Indeed, we noticed that bedload transport died out rapidly  once the flume was no longer supplied with particles. This provides evidence that transport rates were mostly controlled by the feed rate and only weakly by flow strength.

In more realistic situations, $\varepsilon$ is close to, but larger than, $0$ (if $\varepsilon<0$, no equilibrium solution for $\gamma^*$ would exist). In such cases, any perturbation in particle activity needs a very long distance to dissipate. The effect of the upstream boundary conditions can thus be felt far downstream.   A typical measure of this distance is the so-called relaxation length $\ell_\text{sat}$, which was obtained analytically in a previous study \citep{heyman14}. Notably, $\ell_\text{sat}$ grows rapidly as the inverse of $\varepsilon$. Consequently, our results suggest that the transport of coarse particles over a steep slope may depend on transport conditions occurring far upstream. This apparent nonlocal effect may be another explanation for why algebraic bedload transport equations (relating bedload to local flow conditions) fail to accurately predict transport rates. This possibility has been also evoked by \citet{tucker10}, \citet{furbish13b} and \citet{ancey15jgr}.

\section{Conclusions}
Recent stochastic models of bedload transport demand that particle dynamics be described in some detail. These models also need closure equations if one wishes to apply them to nonuniform flow conditions. In particular, how particle diffusivity, entrainment and deposition rates vary with flow conditions is of paramount importance to numerical simulations \citep{bohorquez15}. A single article will not be sufficient to cover all the aspects of closure equations. In this paper, the emphasis is thus placed on the dynamics of bedload transport in one-dimensional shallow supercritical flows on sloping mobile beds. Our setting is representative of flow conditions encountered in mountain streams.

We ran experiments with well-sorted natural gravel in a narrow flume.  Using a fast imaging technique coupled to a particle tracking algorithm, we collected a large sample of particle paths (all in all, more than 8~km of trajectories were reconstructed at the grain scale). At the same time, we measured the evolution of the bed and water profiles, which allowed us to directly relate the particle dynamics to local flow conditions.

On the whole, we found that particle acceleration, velocity, diffusivity and deposition rate were closely associated with local flow and bed conditions. The following closure equations matched our data:
\begin{eqnarray}
\left\langle{u_p}\right\rangle&=&0.57 \bar{u} -0.16 w_s  \quad \left(  \left\langle{u_p}\right\rangle=3.6 u_* -0.06  w_s \right), \label{eq:closure1}\\
u_p/\bar{u} &\sim& \mathcal{N}(0.4,0.3), \\
D_u^*&=&1.6 \bar{u}/w_s - 1.02,  \\
D_p^* &=& 0.0037 \left(\tau^*- \tau_\text{cr}^*\right )^{-1/2},\\
D_p^* &=& 0.05 \tan \theta_k + 0.017 \text{ for } k=10. \label{eq:closure2}
\end{eqnarray}
The equations just above hold for a narrow range of flow conditions: $d\approx6 $mm, $\text{Fr}>1$,  $0.1<u_*/w_s<0.3$, $\tau^*-\tau_\text{cr}<0.25$, and $-0.2<\tan \theta<0.1$. In this paper, we took a first step towards closing stochastic and deterministic non-equilibrium bedload models \citep{charru06,audusse15,bohorquez15,bohorquez15b}. Extending Equations \eqref{eq:closure1}--\eqref{eq:closure2} to a wider range of flow conditions requires much more work.

Surprisingly, the areal entrainment rates $E$ showed only weak dependence on hydraulic and topographic variables, but strong dependence on local particle activity. Specifically, particle entrainment was greatly facilitated by the passage of moving particles and justified the decomposition of $E$ into a flow contribution $e_0$ and a particle activity contribution $e_1$, as proposed by \citet{ancey08a}. Our experiments further suggest that, at low Shields numbers, $e_0 \approx 0$ and $e_1 \approx D_p^*$.

Furthermore, we emphasize several additional findings regarding particle dynamics:
 \begin{itemize}
 \item The traditional view in classical bedload models is that transport capacity tends to zero (or negligibly small values) as the Shields number approaches a threshold (or a reference value). If so, in the absence of significant particle transport in the flume, supplying the flume inlet with sediment should result in bed aggradation. In our experiments, however, neither bed aggradation nor degradation was observed, regardless of the feed rate imposed. This may suggest that transport of coarse grains over steep slope differs a great deal from what is usually observed at shallower slopes and for finer particles \citep{lajeunesse10,houssais15}. This may also reveal a limitation in our experimental set-up. Indeed, we cannot exclude the possibility that the particle kinetic energy initially imparted by the conveyor belt was sufficiently high for the particles to stay in motion during a few hops. Before they came to a halt, these particles destabilized other particles resting on the bed interface, and thereby they initiated low sediment transport.

\item Our findings are in line with recent simulations based on discrete element methods. \citet{maurin15} obtained a similar particle velocity profile to the one reported in Fig~\ref{figure4}a. \citet{Clark2015} showed how important the particle Stokes number is when studying particle dynamics at the onset of motion. Earlier investigations demonstrated that particle collision in a viscous fluid is quasi-elastic when the Stokes number exceeds 1000 \citep{gondret99,joseph01,schmeeckle01}. A likely consequence is that, part of the momentum carried by moving grains is transferred to bed particles, occasionally causing them to be set in motion.
 \end{itemize}

\begin{acknowledgements}
This work was supported by the Swiss National Science Foundation under grant number 200021\_129538 (a project called ``The Stochastic Torrent: stochastic model for bed load transport on steep slope''), by R'Equip grant number 206021\_133831, and by the competence center in Mobile Information and Communication Systems (grant number 5005-67322, MICS project). P.B. acknowledges the financial support from MINECO/FEDER (CGL2015-70736-R), the Caja Rural Provincial de Ja\'en and the University of Ja\'en (UJA2014/07/04). We are grateful to Bob de Graffenried for his technical support. We thank the two anonymous reviewers, Martin Raleigh and John Major,  Associate Editor Amy East, and Editor John M. Buffington, for their constructive remarks and their remarkable review work. Data and Matlab scripts are available online from {https://goo.gl/p4GbsR}.
\end{acknowledgements}
\end{article}
%

\end{document}